# SOME INTERESTING PHENOMENA OBSERVED IN COSMIC-RAY EXPERIMENTS BY MEANS OF X-RAY EMULSION TECHNIQUE AT SUPER ACCELERATOR ENERGIES

*V. M. Maximenko, V. S. Puchkov, S. E. Pyatovsky, S. A. Slavatinsky, and others*

P. N. Lebedev Physical Institute, Russian Academy of Sciences, Moscow

*R. A. Mukhamedshin*

Institute for Nuclear Research, Moscow

In the energy range above the PCR "knee" (5 - 1000 PeV), the emulsion chamber data exhibit some new features which are hard to explain in the framework of the Standard Model. In this survey the results of a systematic quest of some new and exotic types of events which are observed in the *Pamir* emulsion chamber experiment are presented, namely, huge "halo" events, coplanar events, penetrating hadrons with abnormally weak absorption in lead, "Centauro"-type events. Possible theoretical approaches for their explanation are discussed.

## INTRODUCTION

Experiments with cosmic rays at an energy of $E_0 \geq 10^{15}$ eV provide a unique opportunity to explore particle interactions at super accelerator energies and, on the other hand, to study energy spectrum and composition of the primary cosmic rays (PCR) which relates to fundamental astrophysical problems such as origin, acceleration and propagation of the PCR in the Universe. However, experiments in this energy region are rather complicated, the main difficulty being a very low PCR flux due to its rapid fall with increasing energy. For example, the integral PCR flux at energies above $10^{16}$ eV is only 1 particle/(m$^2$ sr y). Since direct measurements by balloon- or satellite-borne instruments are practically impossible, the only way to investigate high-energy cosmic rays is to record extensive air showers (EAS) induced in the atmosphere by the PCR particles via a large number of successive nuclear interactions followed by electromagnetic cascades.

A usage of large-scale X-ray emulsion chambers (XREC) installed at high mountain altitudes enables experimenters to decrease a strong shielding effect of the atmosphere and to increase sensitivity of ground-based experimental setups to the PCR particles and their initial interactions thanks to less number of nuclear interactions contributing to the events being recorded. The advantages of the XREC technique have been realized in a full swing in the *Pamir* experiment which was launched in 1973 at the Pamirs as a collaborative work of several research laboratories from the former USSR and Poland. In 1980, the physicists engaged in the *Pamir* experiment established close and fruitful contacts with their Japanese and Brazil colleagues conducting similar explorations at the Chacaltaya High Altitude Laboratory. Since that time on, a lot of joint projects were carried out that made it possible to increase efficiency of the researches of common interest [1].

## 1. EXPERIMENTAL SETUP AND TECHNIQUE

The *Pamir* XREC (Fig. 1) represents a solid-state track chamber installed at the Pamirs at the altitude of 4400 m a.s.l. An air layer above the chamber constitutes a thick target (600 g/cm$^2$) where nuclear-electromagnetic cascades (NEC) induced by PCR particles take place. A high-energy electromagnetic component of the cascades initiates in the upper lead plates of the chamber, assembled in a stack and called Γ-block, electron-photon cascades (EPhC) which are recorded as dark spots by X-ray films placed between and under lead plates if energy of incident γ-rays and electrons is high enough ($E_\gamma \geq$ 2 - 4 TeV).

High spatial resolution of the X-ray films with two emulsion layers separated by a plastic support of ~200 μm thick makes it possible to determine coordinates, zenith and azimuthal angles of incident particles with accuracies as high as of Δ$x$, $y$ ~ 100 μm, Δ$θ$ ~ 3°, and Δ$φ$ ~ 15°, respectively.

The darkness of a spot is proportional to the density of cascade electron number and thus to the γ-ray energy $E_\gamma$, which can be determined by photometric measurements of the spot accounting for lateral distribution of cascade electrons. The accuracy of individual EPhC energy determination $\sigma(E_\gamma) / E_\gamma$ = 0.2 - 0.3.

The method of energy determination by XREC has been calibrated in accelerator beams of electrons and pions and proved in the experiment on reconstruction of $π^0$ mass in the decay $π^0 \rightarrow 2γ$ by means of measuring the energy of two initial γ-rays and opening angle Φ: $m_{π0} = \Phi\sqrt{E_1 E_2}$

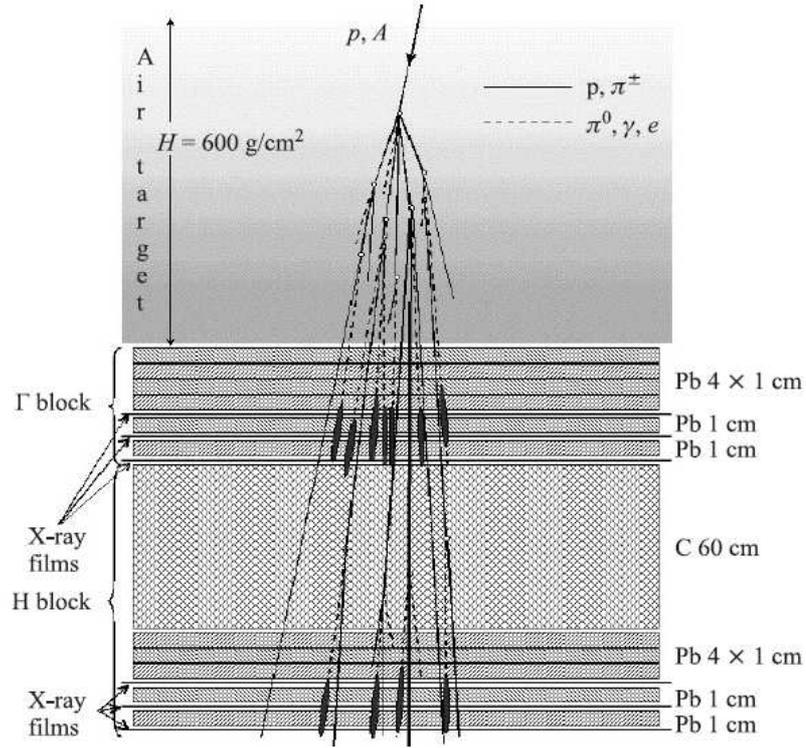

Fig. 1. A lay-out of the *Pamir* experiment setup.

To record high-energy hadrons, two types of XREC are used. The so-called C chamber incorporates a layer of carbon-containing absorber (namely, rubber) under the Γ-block with thickness of about mean free path for interaction in which hadrons interact and transfer a fraction of their energy $E^{(\gamma)}_h = K^{\text{eff}}_\gamma E_h$ into electromagnetic component via decay of produced neutral pions, where effective partial inelasticity coefficient $K^{\text{eff}}_\gamma \sim 0.3$. Since the mean free path for nuclear interaction in lead $\lambda^{\text{Pb}}_{\text{int}}$ exceeds considerably the cascade unit $t^{\text{Pb}}_0$ ($\lambda^{\text{Pb}}_{\text{int}}/t^{\text{Pb}}_0 \approx 30$), the EPCs induced by γ-rays prevail in the Γ-block, while hadrons predominantly interact and are recorded deeper in the XREC, i.e., in H-block. For energy determination of both gammas $E_\gamma$ and hadrons $E^{(\gamma)}_h$ recorded in C chamber, only one layer of X-ray films at the fixed depth $t = 8 - 12$ c.u. is used, which corresponds approximately to the position of the EPhC maximum in a wide energy range (one-point method).

Another basic type of the *Pamir* XREC for hadron recording is the so-called Pb chamber, i.e., deep uniform lead XREC assembled of lead plates only. The upper part of Pb chamber is similar to Γ-block, but the lower part contains at least 40 - 60 plates of lead (1 cm thick each) interposed with X-ray films, since determination of hadron energy $E^{(\gamma)}_h$ released in lead requires darkness measurements at several depths to find the whole cascade transition curve (multipoint method).

The *Pamir* XREC contains 1500 t of rolled lead plates and 620 t of carbon-containing materials that defines its unique features, namely, large area (~ 1000 m²) and high efficiency of hadron detection (60 - 75 %). The instrument is deployed inside a shed which is equipped with electrical machinery and vacuum pump system used for chamber assembling works. For its operation, the setup annually requires 4000 - 7000 m$^2$ of high sensitive X-ray films of RT-6 type and its maintaining is confined to annual change of exposed films by new ones. Processing and analysis of experimental material is carried out at laboratory conditions employing photodensitometers and scanners.

Due to high-energy threshold for particle detection and high spatial resolution, XRECs make it possible to resolve narrow collimated bundles of the most energetic particles in the central part, i.e., core, of the corresponding EAS and thus to study semi-inclusive spectra of secondary particles practically in the whole fragmentation region of the projectile. These bundles of high-energy $\gamma$-rays and hadrons are recorded in X-ray films as groups of darkened spots and are called families.

The quantitative criteria for $\gamma$-family selection are the following: 1) number of $\gamma$-rays $n_\gamma \geq 3$; 2) $E_\gamma \geq E_{th} = 4$ TeV; 3) incidence directions of all $\gamma$-rays in a family coincide at least within accuracy of angle measurement, i.e., $\Delta\theta = 3°$, $\Delta\varphi = 15°$; 4) distance of an individual $\gamma$-ray from the energy-weighted centre of family at the target plot $R_{\gamma i} \leq R_{max} = 15$ cm; 5) total energy of a family $\Sigma E_{\gamma i} \geq 100$ TeV.

Since the air target above the XREC is thick enough, several generations of NEC contribute to the observed families that makes analysis of experimental events rather complicated and requires a detailed simulation of the experiment including that of the chamber response. Experimental data are compared with the model simulations based on different versions of the quark-gluon string (QGS) model which provides a nice fit of high-energy data in a wide accelerator energy range. In practice, we use very efficient MQ [2] and MC0 codes [3] which are close to the CORSIKA one based on the QGSJet model accounting for hard QCD-jet production. Simulation of the chamber response including conditions of experimental event selection and final spatial resolution of particles are correctly incorporated into the model sampling.

In the energy range $E_0 = 10^{15}$ - $10^{16}$ eV, the simulations fit the experimental data very well [4], but there are significant discrepancies in the range $10^{16}$ - $10^{17}$ eV. The most challenging phenomena are: a) a high intensity of *halo* events; b) the coplanar emission of the most energetic hadrons and $\gamma$-rays in the multiple particle production; c) Centauro events with abnormal ratio of charged and neutral particles; d) abnormal behavior of a hadron absorption curve, which significantly deviates from exponential law at large depth in lead absorber.

## 2. HALO EVENTS

If the energy $E_0$ of a PCR particle is about or higher than $10^{16}$ eV, a sufficiently high number of overlapping under-the-threshold EPhC may overlap creating an optical "halo", i.e., a large diffuse dark spot (Fig. 2, *a*) inside the corresponding $\gamma$-family with a visible energy $\Sigma E_\gamma \geq 500$ TeV. Sometimes area $S$ of a halo amounts to several centimeters squared. The fraction of halo events increases with family energy and, at $\Sigma E_\gamma \geq 1000$ TeV, amounts up to 70 %. Previous simulations revealed that a halo may be induced by a pure EPhC initiated by a neutral pion in the atmosphere or by a narrow bundle of many EPhC directly related to a PCR particle nuclear interaction occurring not so far above the observation level.

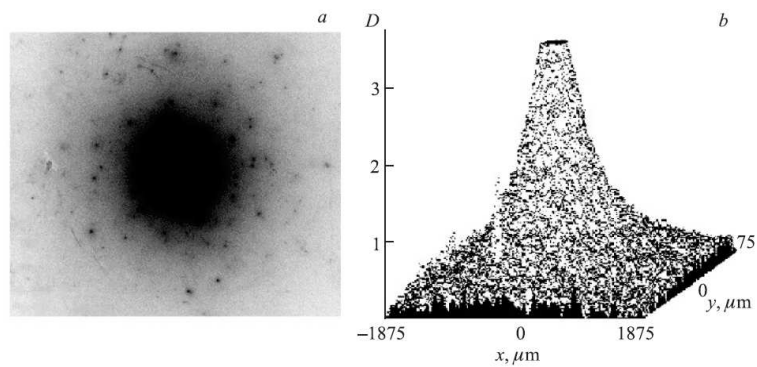

Fig. 2. Examples of halo phenomenon observation and halo processing: *a*) a scanner image of a huge halo accompanied by high-energy $\gamma$-rays which belong to one of the most energetic events observed in the *Pamir* experiment and called "FIANIT" with visible energy $\Sigma E_\gamma \approx 20$ PeV and estimated primary energy $E_0 \sim 4 \times 10^{17}$ eV; *b*) a 3-dimensional densitogram of a halo with energy 5 PeV and area $S$ equal to 6 mm$^2$ bounded by isodense with $D = 0.5$.

We analyze 143 super-families with $\Sigma E_\gamma \geq 500$ TeV (including 60 halo events) recorded in the *Pamir* experiment during exposure $ST \sim 3000$ m$^2$ y of all types of XRECs containing Γ-block. Densitograms of halos (Fig. 2, *b*) were measured in X-ray films exposed at the depth of 5 cm of lead, i.e., 9 - 11 c.u. taking into account the incidence angle. We accept the following criteria of halo event selection: family energy $\Sigma E_\gamma \geq 500$ TeV, the area $S$ bounded by the isodence with optical density $D = 0.5$ (particle number density 0.04 $\mu$m$^{-2}$) above the background at the depth of 5 cm of lead exceeds 4 mm$^2$ or, in the case of multi-core (structural) halo $\Sigma S_i \geq 4$ mm$^2$, where area of each halo core $S_i \geq 1$ mm$^2$.

*Table* 1. The fraction of multi-core halo $\gamma$-families produced by different primary particles.

| *p* | *α* | C | Fe | MC0 | Pamir |
|---|---|---|---|---|---|
| 0.25 ± 0.03 | 0.45 ± 0.09 | 0.59 ± 0.11 | 0.70 ± 0.12 | 0.28 ± 0.03 | 0.23 ± 0.07 |

A new set of family simulations at a wide energy range $2 \times 10^{15}$ - $3 \times 10^{18}$ eV have been recently performed in current MC0 and MQ codes accounting for halo creation. The PCR particles with isotropic angular distribution at the atmospheric boundary are sampled by Nikolsky's spectrum [5] with a single exponent of integral spectrum $\gamma = 2.05$. The PCR absolute intensity is taken as $N(E_0 > 10 \text{ PeV}) = 3 \times 10^8$ m$^{-2}$ s$^{-1}$ sr$^{-1}$. The zenith angles vary from 0 to 50°. Conventional or "normal" mass composition of the PCR particles is used. It means that fraction of protons and helium nuclei in the PCR is about 50 % at an energy $E_0 = 1$ PeV and slowly decreases with energy.

All particles of a NEC in the atmosphere are traced down to an energy of 100 GeV. Simulation of the halo formation in X-ray films is performed using the so-called "isodense" method described in [6]. The calculations reveal that contribution of particles with energy less than 100 GeV to halo areas is about (3 - 5) % for small halo areas ($S = 4 - 10$ mm$^2$) and near (10 - 15) % for large halos ($S \geq 100$ mm$^2$). This effect is taken into account. The total statistics of the simulated atmospheric cascades amounts 40 thousand events and 800 halo families are among them.

According to the calculations, the majority of halo events are produced by the PCR particles with energies in the range of $E_0 = 5 \times 10^{15} - 10^{18}$ eV. The probabilities of halo event production by different nuclei, calculated in several model codes, confirm that, at an energy $E_0 \sim 10^{16}$ eV, protons produce halo events several times more efficiently than other nuclei (29 % of primary protons produce 82 % of total number of halo events). At a higher PCR energy ($E_0 \sim 10^{18}$ eV), the probability of production of $\gamma$-families with halo approaches unity for all nuclei, but contribution of the PCR particles with such an energy to the overall flux of halo events is negligible. Hence, the intensity of halo events and the spectrum of halo areas are sensitive mainly to the fraction of protons in the PCR flux. A detailed analysis of the dependence of halo event fraction $N_{halo}/N_{tot}$ in the interval $\Sigma E_\gamma = 500 - 1000$ TeV on atomic number $A$ of the PCR particles makes it possible to conclude that the experimental data are satisfactorily fitted by the MCO code and, at energy $E_0 \sim 10^{16}$ eV, the average $<\ln A>$ is not more than 2.5.

The calculations also show that primary protons produce predominantly one- or two-core halos while nuclei create multi-core halos ($N_c \geq 2$). As is seen from Table 1, experimental data on multi-core halo event fraction are also well fitted by the MCO model. The comparison confirms that the majority of halo events are produced by the primary protons.

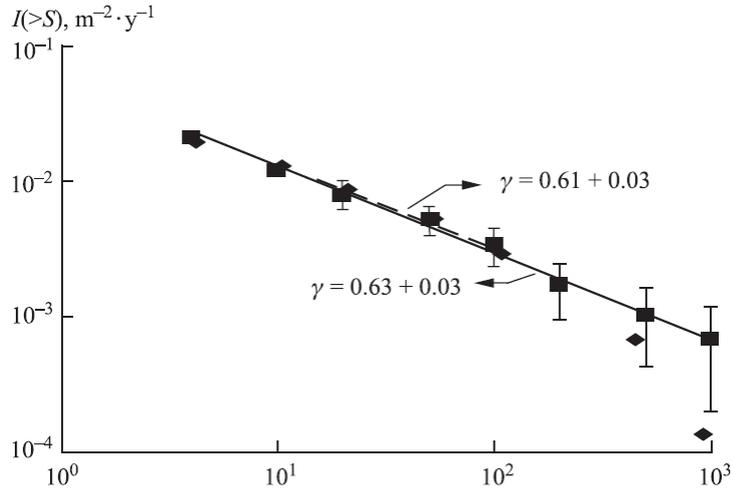

Fig. 3. Experimental and simulated integral distribution of halo areas: ♦, dashed line - MC0; ■, solid line - experiment.

As is seen in Fig. 3, experimental spectrum of halo areas and absolute intensity of halo events are consistent with simulated ones in the area range $S = 4 - 200$ mm$^2$ ($E_0 \approx 5\times10^{15} - 10^{17}$ eV) under assumption that proton fraction in the PCR is close to 25 % at $E_0 \sim 10^{16}$ eV and the PCR mass composition is gradually enriched with heavy nuclei. But, at $S \geq 200$ mm$^2$, the calculated spectrum becomes appreciably steeper than the experimental one, and, at $S \sim 1000$ mm$^2$, the difference is as large as 4 - 5 times. To match the experiment and calculations, it is necessary to assume that, at $E_0 = 10^{17} - 3\times10^{18}$ eV, the proton fraction in the PCR composition gradually increases up to 100 % while the PCR intensity at $E_0 = 10^{18}$ becomes 2 - 3 times larger. Such a growth of the PCR intensity may be explained assuming that exponent of the PCR differential energy spectrum at $E_0 \geq 10^{17}$ eV regains the same value (2.6 - 2.7) it has at energies above the "knee" of the PCR energy spectrum.

The third possibility to match the experiment and calculations is based on a hypothesis of penetrating particles in the PCR composition. These particles may interact (or decay) close to the observation level and produce narrow bundles of $\gamma$-rays and hadrons which can give rise to large halo formation in the XREC.

### 3. PHENOMENON OF COPLANAR EMISSION OF HADRONS

The *Pamir* experiment revealed [7] that, besides single-core *halos*, there exist *halos* with 2, 3, 4 and more number $N_c$ of cores. It is remarkable that core centers of multi-core *halo* events with $N_c \geq 3$ exhibit a strong tendency to be aligned along one straight line in the target plane. Up to now, we have analyzed 9 super-families with $N_c \geq 3$ and 6 of them appeared to be aligned. Soon it was established that this effect manifests itself in the highest energy region ($\Sigma E_\gamma \geq 700$ TeV) through alignment of various most energetic cores inside $\gamma$-hadron families represented not only by halo cores but also by the highest energy $\gamma$-rays, hadrons and $\gamma$-hadron clusters [8, 9]. The treating of the highest energy cores (HEC) in families makes it possible to analyze any γ-hadron family and therefore to increase the experimental statistics.

The similar but single events were observed in some other cosmic ray experiments with XREC exposed both at mountain elevation [10] and in the stratosphere [11, 12]. Some years ago Chinese physicists, who had accomplished in the 90's the *Mt. Kanbala* experiment at the height of 5600 m a.s.l., reprocessed their experimental data, amounting a few dozens of super-high energy events recorded with iron XREC, and confirmed the main conclusions of the *Pamir* experiment within available statistics [13].

For quantitative definition of events with $N$ cores aligned along one straight line, the following criterion was introduced:

$$\lambda_N = \frac{\sum\limits_{i \neq j \neq k}^{N} \cos 2\varphi_{ij}^k}{N(N-1)(N-2)}$$

where $\varphi^k_{ij}$ is the angle between the straight lines connecting the $i^{th}$ and $j^{th}$ cores with the $k^{th}$ core. The parameter $\lambda_N = 1$ in the case of complete alignment of $N$ cores along one straight line and tends to $-1/(N-1)$ in an isotropic distribution case. Families containing $N$-core structures, composed of the HEC and characterized by $\lambda_N \geq \lambda_c = 0.8$, are referred to as aligned events. The critical value $\lambda_c = 0.8$ was chosen according to comparison of experimental $\lambda_N$ distributions for 3, 4, 5 and 6 HEC with model simulations taking into account statistical fluctuations and correlations due to propagation of NEC through the atmosphere. These simulations show that contribution of statistical fluctuations and conventional processes to that particular $\lambda_N$ range is very low [8] and presents a background of the effect.

We can restrict the study with analyzing only γ-families to avoid a problem of unification of reference frames used for gamma and hadron treating and thus to use the whole available experimental statistics. However, this approach may decrease sensitivity of extracted HEC to the effect under consideration due to discrimination of hadron component. The total number of γ-families with $\Sigma E_\gamma \geq 100$ TeV ($E^{(i)}_\gamma \geq 4$ TeV, $R^{(i)}_\gamma \geq 15$ cm) observed in various types of XREC is 974 events and 62 of them are in the energy range $\Sigma E_\gamma = 700 - 2000$ TeV.

To reduce a destructive effect of NEC propagation on observation of coplanar emission of hadrons by XREC at mountain altitudes, we apply consecutively three procedures of γ-family "rejuvenation" which make it possible to treat elder particles. The first of them reconstructs initial γ-rays by means of electromagnetic decascading algorithm [14] which employs parameter $z_{ik} = r_{ik}(1/E_i + 1/E_k)^{-1}$, where $E_i$ is the energy of unified gammas and $r_{ik}$ is the distance between $i^{th}$ and $k^{th}$ particles. All closely spaced particles with $z_{ik} \leq z_c = 1.2$ TeV cm are combined in one electromagnetic cascade induced by the corresponding initial γ-ray. The critical value of $z_c$ is derived from the model simulations and has a clear physics meaning since, according to Coulomb multiple scattering theory, about one half of cascade particles with energy $E$ are within the radius $r_c \sim (21 \text{ MeV}/E) t \approx 1$ mm at the Pamirs height, where the cascade unit length $t = 500$ m.

A dissection of simulated $\gamma$-families calculated in MC0 code reveals that the same $z_{ik}$-algorithm can be used for coupling pairs of initial $\gamma$-rays and reconstruction of corresponding $\pi^0$ produced in the last NEC interactions just above the chamber. At $z_c = 3.4$ TeV cm, the efficiency of $\pi^0$ reconstruction is close to 90 %.

A nuclear clustering procedure [15] based on $\chi_{ik}$ algorithm, i.e., $\chi_{ik} = \sqrt{E_i E_k} \, r_{ik} \leq \chi_c = 48$ TeV cm, effectively combines particles arriving from individual nuclear interactions which occur just above the chamber. The extracted hadronic clusters practically represent individual hadrons ($h^*$) from the last but one NEC generation. The efficiency of $h^*$ reconstruction in $\gamma$-families with $\Sigma E_\gamma = 100 - 2000$ TeV ($E_\gamma > 4$ TeV, $R_\gamma \leq 15$ cm) is equal to $(81 \pm 1)$ %. Note that $\chi_c$ value has also a clear physics meaning, namely, $\chi_c \approx H p_t$, where $H \approx 1.2$ km is the hadron interaction height above the chamber and $p_t \approx 0.4$ GeV/$c$ is the mean transverse momentum in strong interactions.

The performed analysis of experimental "rejuvenated" $\gamma$-families shows that, at the low $\gamma$-family energy range $\Sigma E_\gamma = 100 - 400$ TeV, the probability of observation of aligned 4-core structures, consisting of either the highest energy $\gamma$-rays, neutral pions or $h^*$-clusters, is well fitted by model calculations and application of the above-described decascading (clustering) procedures does not influence the aligned event fraction. However, at higher energies ($\Sigma E_\gamma \geq 700$ TeV), we observe a considerable increase of efficiency for aligned HEC observation at the physically meaningful values of $z_c$ and $\chi_c$ parameters that strongly favors a hypothesis of coplanar emission of high-energy pions at super-high energies. In particular, the experimental fraction of aligned super-high energy events with 4-core structures composed of reconstructed $\pi^0$ is as high as $F^{\text{exp}}_\pi (\lambda_4 \geq 0.8) = 0.22 \pm 0.05$ that exceeds the background value $F^{\text{MC0}}_\pi (\lambda_4 \geq 0.8) = 0.06 \pm 0.01$ by $\sim 3$ standard deviations.

The fraction of $\gamma$-families with aligned 4-centre HEC structures is especially high and amounts, in the case of hadronic clusters, $F^{\text{exp}}_h (\lambda_4 > 0.8) = 0.27 \pm 0.09$, if one selects in advance only families containing not less than 6 hadronic clusters ($N_c \geq 6$) with energy $E_c \geq 50$ TeV (see Table 2). Moreover, for families with $N_c < 6$ ($E^{\text{th}}_c = 50$ TeV) the corresponding value is near to the background, the value of which ($\sim 4$ %) is independent of $N_4$ (Fig. 4).

*Table* 2. Fractions *F* of aligned 4-core structures composed of reconstructed particles and their mean sizes $<R_c>_4$ or γ-families selected under a condition $N_c > 6$ ($E^{th}_c = 50$ TeV)

| Type of events | Examined parameters | Type of reconstructed particle | | |
|---|---|---|---|---|
| | | γ* ($z_c$ = 1.2 TeV cm) | $π^{0*}$ ($z_c$ = 3.4 TeV cm) | h* ($χ_c$ = 48 TeV cm) |
| EXP | $F(λ_4 > 0.8)$ | 0.13 ± 0.04 | 0.22 ± 0.05 | 0.27 ± 0.09 |
| | $<R_c>_4$, cm | 1.5 ± 0.5 | 1.3 ± 0.3 | 1.8 ± 0.5 |
| MCCP | $F(λ_4 > 0.8)$ | 0.16 ± 0.01 | 0.24 ± 0.01 | 0.20 ± 0.02 |
| | $<R_c>_4$, cm | 1.2 ± 0.1 | 1.4 ± 0.1 | 1.2 ± 0.1 |
| MC0 | $F(λ_4 > 0.8)$ | 0.06 ± 0.01 | 0.07 ± 0.01 | 0.05 ± 0.01 |
| | $<R_c>_4$, cm | 0.6 ± 0.1 | 0.8 ± 0.2 | 0.6 ± 0.1 |

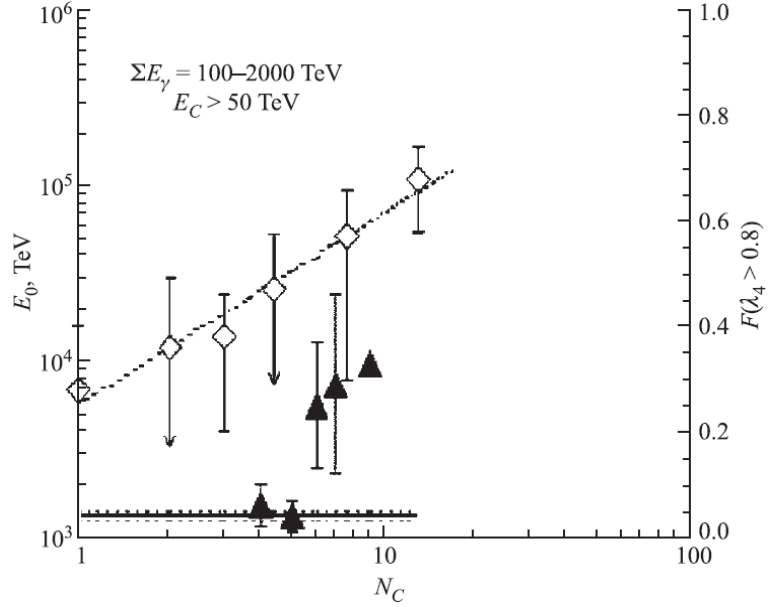

Fig. 4. Dependence of primary cosmic ray particle energy $E_0$ and fraction *F* of γ-families with aligned 4-core cluster structures on total number $N_c$ of extracted nuclear clusters with $E_c \geq 50$ TeV: ◊ - $E_0$: exp; ▲ - *F*: exp; solid line - *F*: MC0.

The last result is easy to conceive assuming an energy threshold behavior of the phenomenon under consideration. Unlike a weak dependence of the PCR particle energy $E_0$ on the visible γ-family energy $ΣE_γ$, there is a strong correlation between the number $N_c$ of hadronic clusters extracted in an observed family and the energy $E_0$ of the corresponding PCR particle, particularly that of the primary proton (Fig. 4). Referring to Fig. 4, it is possible to estimate at 68 % confidence level the energy threshold of the coplanar event production by the primary protons, namely, $E^{th}_0 \geq 8$ PeV.

The comparison of aligned ($λ^c_4 > 0.8$) experimental γ-families with nonaligned ($λ^c_4 < 0.8$) ones selected under the condition $N_c \geq 6$ ($E_c > 50$ TeV), makes it possible to conclude that their overall or integral characteristics are practically the same. The difference appears when we treat the HEC in the central part of a family, namely, the lateral spread of four most energetic clusters. The average radius of experimental aligned structures $<R^4_c>^{exp} = 1.8 ± 0.5$ is about three times higher than that given by MC0-model simulation, i.e., $<R^4_c>^{MC0} = 0.6 ± 0.1$ which is in agreement with experimental value for nonaligned events. This result, on the one hand, favors a large transverse momentum of the coplanar emission process, and, on the other hand, testifies to strong destructive effect of a subsequent or/and accompanying NEC on initial coplanar pattern. A ratio of longitudinal component $\langle p_t \rangle^{\parallel}$ of the average transverse momentum of secondaries to transverse component $\langle p_t \rangle^{+}$, determined in reference to the coplanarity plane, can be estimated by the use of corresponding scattering of aligned cores in experimental families with respect to the line of their alignment, i.e., $\langle p_t \rangle^{\parallel}/\langle p_t \rangle^{+} \approx \langle R_c \rangle^{\parallel}_4 / \langle R_c \rangle^{+}_4 = 12 \pm 3$. If one assumes, that transverse component $\langle p_t \rangle^{+}$ of the process corresponds to the mean transverse momentum of hadrons inside QCD jets, i.e., 0.22 GeV/$c$, then the transverse momentum of the process of coplanar emission of the highest energy hadrons may amount several GeV.

As regards the attenuation of the alignment effect with NEC propagation, this conclusion is consistent with another distinguishing feature of aligned experimental $\gamma$-families as compared with nonaligned ones, i.e., a significant difference in the mean zenith angle $\theta$ of primary particles, namely, $\langle \cos \theta \rangle = 0.968 \pm 0.009$ for aligned events ($\lambda^c_4 > 0.8$); while for nonaligned ones, $\langle \cos \theta \rangle = 0.923 \pm 0.015$. That means that inclined showers do not effectively preserve the aligned structures due to greater number of contributing successive interactions.

To determine the destruction rate of a coplanar pattern of secondary hadrons produced at the top of the atmosphere by their successive interactions with air nuclei above the chamber, a phenomenological model of coplanar particle production was incorporated in the MC0 code [16]. The developed MCCP model assumes a step-like threshold for coplanar particle production in the $p$-$A$ interaction at an energy of $E^{th}_0 = 8$ PeV. At energies above the threshold, the strong interactions drastically change since the probability for the process of coplanar emission of secondaries at these energies was taken to be equal to 1. The coplanarity of the particle emission means that longitudinal component of the average transverse momentum of secondaries in reference to the coplanarity plane $\langle p_t \rangle^{\parallel}$ is as high as 2.3 GeV/$c$ while its transverse component $\langle p_t \rangle^{+} = 0.22$ GeV/$c$.

As is seen from Table 2, which contains fractions of aligned 4-core structures composed of reconstructed particles and their mean sizes, the model provides a good fit to experimental data on the alignment phenomenon.

It is quite evident that conventional hard interaction processes could not completely vanish. To make inelastic cross section of coplanar particle production realistic, one needs to introduce also penetrating particles in order to decrease the destruction of coplanar frames by NEC. A mean free path of 200 - 300 g/cm$^2$ for a penetrating particle, producing coplanar hadrons in nuclear interactions, seems to be optimal to accord all variety of observable characteristics.

Most of hypotheses on the aligned event production mechanism in the framework of QCD were considered in [16]. One of them is based on the one-dimensional character of gluon strings which rapture produces aligned images of emitted particles in films [17]. However, this simple picture does not agree with experimental data since, in the experimental case, particles responsible for the alignment effect are the most energetic ones in the parent interaction. These particles are produced in the fragmentation region while semi-hard strings are produced in the pionization, i.e., central part region of strong interaction. Another possible mechanism considered in [18] assumes the aligned particle production in the result of inelastic double diffraction. This hypothesis made it possible to understand some important peculiarities of the alignment effect, namely: a) a large fraction of energy carried by aligned particles, b) high probability of production of 4, 5, etc., aligned HECs, c) existence of the energy threshold for aligned event production of about 10 PeV, and, at last, d) the high value of the $\langle p_t \rangle^{||} / \langle p_t \rangle^{+}$ which equals ~ 10. However, according to theoretical concepts, the transverse cross section of inelastic double diffraction is rather small and amounts about 10 % at ~ 700 PeV.

One more attempt to explain the aligned event production [19] assumes production of a new type of particles like high color hadron sextets at EeV energy region. Although such an exotic hypothesis makes it possible to explain high efficiency of aligned event production, it seems to be rather premature as concerns to the Standard Model restrictions.

## 4. CENTAURO EVENTS

In 1972 the Japan-Brazil Collaboration observed an unusual event in a two-storied emulsion chamber exposed at the Mt. Chacaltaya at a height of 5200 m a.s.l. This event consisting of 49 hadrons with total energy of 230 TeV and containing nothing released in electromagnetic component was named a Centauro event [20]. It proved to be difficult to explain such an event by a fluctuation in the production ratio of charged to neutral pions since its probability is negligible [21].

*Table* 3. Characteristics of Centauro-type events

| Experiment | $\Sigma_\gamma + \Sigma E^\gamma_h$, TeV | $n_\gamma + n_h$ | Energy threshold |
|---|---|---|---|
| Mt. Chacaltaya | ? + 230 <br> 24 + 179 <br> 101 + 168 <br> 139 + 148 <br> 79 + 270 <br> 0 + 51 | ? + 49 <br> 5 + 32 <br> 26 + 37 <br> 68 + 39 <br> 25 + 40 <br> 0 + 13 | ~ 1 TeV |
| Pamir | 53 + 209 <br> 600 + 1100 | 6 + 12 <br> 78 + 23 | ~ 4 TeV |

For the next 30 years, no more pure Centauro events with so large number of charged pions unaccompanied by neutral pions were detected not only in the Mt. Chacaltaya emulsion experiment but in other XREC experiments, as well. Nevertheless, a number of Centauro-like events with abnormally high fraction of energy carried by charged hadrons as compared to that of gammas have been observed.

A systematic quest of Centauro-type events was carried out also in the *Pamir* experiment. We analyzed 88 $\gamma$-hadron families with visible energy greater than 100 TeV which were detected by Pb chambers 60 cm thick with total exposition 132 m$^2$ y. Only those families which possessed a large fraction of energy $Q_h$ released in hadrons, namely, $Q_h = \Sigma E^{(\gamma)}_h / (\Sigma E_\gamma + \Sigma E^{(\gamma)}_h) > 0.5$, were examined in detail. An absorption behavior of all particles in such a family was carefully analyzed in order of proper separation of hadrons and gammas. Most of the events have a rather small number of hadrons. Only two of them contain more than 11 hadrons with energy above the threshold $E^{(\gamma)}_{th} = 4$ TeV.

Table 3 shows a comparison of characteristics of Centauro-type events detected in the *Mt. Chacaltaya* and the *Pamir* experiments. A $n_h$ - $Q_h$ scatter plot (Fig. 5) exhibits a satisfactory agreement between experimental and simulated data in a whole area except for the region with $n_h \geq 12$ and $Q_h \geq 0.7$ which contains all Centauro-type events.

As known, a search for Centauro-type events was undertaken in the UA-5 accelerator experiment at CERN. No Centauro-type events were observed among 48000 detected events. The upper limit for production cross section of Centauros was estimated to be $\sigma^{prod}_C < 1\ \mu$b. However, one can argue this negative result that the energy available at CERN was insufficient for Centauro-event production. Another plausible explanation of the negative result is an inadequate accounting for fluctuations in simulations. We cannot also exclude the possibility of Centauro events being of astrophysical origin, and then cosmic rays may be the only way to observe them.

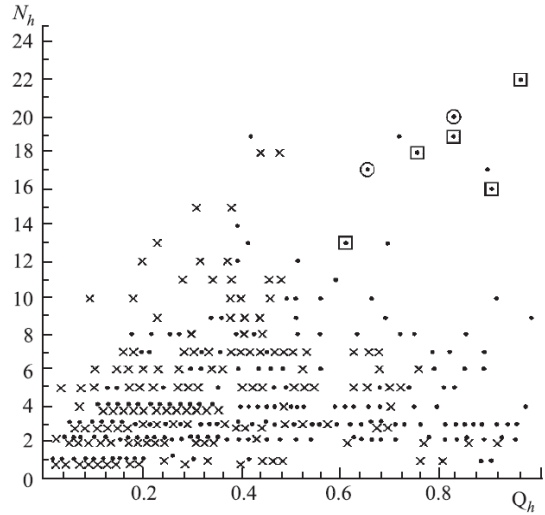

Fig. 5. Correlation plot: number of hadrons $N_h$ versus portion of hadron energy $Q_h = \Sigma E^{(\gamma)}_h / \Sigma E_{tot}$. Dots stand for experimental events while crosses refer to simulated ones. Candidates for Centauro-type events in the *Pamir* and the *Chacaltaya* experiment data are marked by squares and circles, respectively.

## 5. PENETRATING HADRONS WITH ABNORMAL ABSORPTION

One more unusual phenomenon is observed in the *Pamir* experiment which seems to be similar to that discovered earlier at the Tien Shan Mountain Station when the absorption of EAS hadron cores in the hadron calorimeter was studied. It proved to be rather weak as compared to the conventional nuclear one and, to explain the experimental results, a hypothesis of the so-called long-flying component of cosmic rays was introduced [22]. Concerning the *Pamir* experiment, an absorption curve for hadrons with energy greater than 6.3 TeV was obtained by means of deep Pb chambers 120 cm thick. In a range of 0 - 70 radiation lengths, the absorption curve obeys the standard exponential law with index $\lambda_1 = (200 \pm 5)$ g/cm$^2$. However, at larger depths (> 70 c.u.), the absorption length of hadrons in lead changes and becomes as high as $\lambda_2 = (340 \pm 80)$ g/cm$^2$ [23]. It is natural to assume that the excess of hadrons at larger depths is due to a high fraction of charmed $\Lambda_c$ hyperons and *D* mesons produced in the upper lead layers of Pb chamber. Indeed, the decay path of charmed particles with energy $\geq 6.3$ TeV is about 1 m, which is comparable with the chamber depth, and a noticeable fraction of them experienced lepton decays while their collision path is rather large. However, simulation calculations show that the excess of simulated hadron cascades occurring at large depths is in good agreement with the experimental value if one assumes that the production cross section of charmed particles is as high as $\sigma^{prod}_{\Lambda c,D} \approx 2$ mb/nucleon, while extrapolation of accelerator data to the energy range under investigation gives one order less value for $\sigma^{prod}_{\Lambda c,D}$.

In order to check this interpretation of the experimental data, a special experiment with two-storied emulsion chamber is currently running at the Pamirs. This chamber has a 2.5 m air gap where an effective decay of charmed particles should take place resulting in occurrence of a bump of electromagnetic origin at the beginning of the absorption curve.

## CONCLUSIONS

1. While observation of high energy PCR with $E_0 = 10^{15} - 10^{18}$ eV by XREC deployed at mountain altitudes, a new interesting class of events, i.e., $\gamma$-hadron families with halo are recorded which proved to be very sensitive to the nature of the primary particle.

At the primary energy of $E_0 \sim 10^{16}$ eV, the fraction of protons in the PCR is $(25 \pm 4)$ % and <ln A> is not greater than 2.5 assuming conventional behavior of the PCR energy spectrum. This *Pamir* experiment result is in contradiction with some EAS experiments and a now popular view that protons disappear from the PCR flux at these energies.

At the range of the highest energies attainable in the *Pamir* experiment, i.e., $E_0 = 10^{17} - 10^{18}$ eV, a considerable excess of events is observed which number is 5 - 7 times higher than the predicted one. Such an excess can be accounted for by increase of the proton fraction in the given energy range which has to be accompanied by simultaneous flatting of the energy spectrum slope up to power index value $\gamma \approx 2.7$. However, a hypothesis of production of new highly penetrating particles is also relevant.

2. A new phenomenon of coplanar emission of the highest energy hadrons in nuclear interactions at super-high energies $E_0 \approx 10^{16}$ eV is observed in the *Pamir* experiment. This phenomenon manifests itself through appearance of aligned $\gamma$-families with visible energy $\Sigma E_\gamma \geq 700$ TeV and is characterized by a sharp energy threshold at $E^{th}_0 \geq 8$ PeV and a large transverse momentum of about several GeV/*c*.

This phenomenon is hard to explain in the framework of the Standard Model of particle interactions.

3. A systematic search for Centauro-type events, which were first observed by Brazil-Japan emulsion chamber collaboration, was carried out in the *Pamir* experiment. Several events with abnormally low ratio of neutral component energy to that of charge component were found. Production of such events favors the hypothesis of Centauro-like interactions which has still no adequate theoretical description.

4. Abnormal absorption of hundreds TeV energy hadrons is observed in the *Pamir* experiment. This effect can be related to abnormally high cross section of charmed-particles production at TeV energy range. In order to prove this hypothesis, a special experiment at the Pamirs and Tien Shan is launched.

5. All unusual events and phenomena observed in the *Pamir* experiment can be accounted for by an assumption of presence of highly penetrating particles in cosmic rays which are able to penetrate deep in the atmosphere and then interact (or decay) nearby XREC producing halo events, coplanar events or Centauro-type events.

REFERENCES

1. *Baradzei L. T. et al. (Chacaltaya and Pamir collab.)* // Nucl. Phys. B. 1992. V. 370. P. 365.

2. *Dunaevsky A.M. et al.* // Proc. of 3rd Intern. Symp. on Very High Energy Cosmic Ray Interactions, Tokyo, 1984. P. 178.

3. *Fedorova G. F., Mukhamedshin R. A.* // Bull. Soc. Sci. Lett. Lodz. Ser. Rech. Def. 1994. V.XVI. P. 137.

4. *Slavatinsky S. A.* // Nucl. Phys. B (Proc. Suppl.). 1997. V. 52. P. 56.

5. *Nikolsky S. I.* // Proc. of 3rd Intern. Symp. on Very High Energy Cosmic Ray Interactions, Tokyo, 1984. P. 507.

6. *Dunaevsky A.M., Pluta H.* // Proc. of 21st Intern. Cosmic Ray Conf. 1989. V. 2. P. 274.

7. *Baradzei L. T. et al.* // Proc. of 3rd Intern. Symp. on Very High Energy Cosmic Ray Interactions, Tokyo, 1984. P. 136.

8. *Baradzei L. T. et al.* // Izv. Akad. Nauk SSSR. Ser. Fiz. 1986. V. 50. P. 2125.

9. *Ivanenko I. P. et al.* // Pis'ma ZhETF. 1992. V. 50, No. 11. P. 2125.

10. *Amato N.M. et al.* Preprint CBPF-NF-056/86. Rio de Janeiro, 1986.

11. *Capdevielle J.-N. et al.* // Proc. of 25th Intern. Cosmic Ray Conf., Durban, 1997. V. 6. P. 57.

12. *Kotelnikov K. A. et al.* // Proc. of 27th Intern. Cosmic Ray Conf., Hamburg, 2001. V. 4. P. 1426.

13. *Xue L. et al.* // Proc. of 26th Intern. Cosmic Ray Conf., Utah, 1999. V. 1. P. 127.

14. *Bayburina S.G. et al.* // Trudy FIAN (Proc. of Lebedev Physical Institute). M., 1984. P. 1-128.

15. *Borisov A. S. et al.* // Nucl. Phys. B (Proc. Suppl.). 1999. V. 75A. P. 144.

16. *Mukhamedshin R. A.* // Nucl. Phys. B (Proc. Suppl.). 2001. V. 97. P. 122.

17. *Halzen F., Morris D.* // Phys. Rev. D. 1991. V. 42, No. 5. P. 1435.

18. *Royzen I. I.* // Mod. Phys. Lett. A. 1994. V. 9, No. 38. P. 3517.

19. *White A. R.* // Proc. of 8th Intern. Symp. on Very High Energy Cosmic Ray Interactions, Tokyo, 1994. P. 468.

20. *Lattes C.M.G. et al.* // Phys. Rep. 1980. V. 65. P. 151.

21. *Hasegawa S. for the Brazil-Japan collab.* ICR-Report-151-87-5. 1987.

22. *Yakovlev V. I.* // Proc. of 7th Intern. Symp. on Very High Energy Cosmic Ray Interactions, Ann-Arbor, 1992. P. 276; 607.

23. *Rakobolskaya I. V. et al.* Peculiarities of Interactions of Superhigh-Energy Cosmic-Ray Hadrons. M.: Moscow State University, 2000 (in Russian).